\documentclass[11pt]{article}
\usepackage{amsmath,amsfonts}
\usepackage{verbatim}
\usepackage{relsize}
\usepackage{enumerate}
\usepackage[inline]{enumitem}
\normalbaselines
\newtheorem{teor}{Theorem}
\newtheorem{prop}[teor]{Proposition}

\newtheorem{coro}[teor]{Corollary}
\newtheorem{rem}[teor]{Remark}

\newenvironment{demo}{\rm \trivlist \item[\hskip \labelsep{\it
      Proof}.]}{\nopagebreak \hfill $\square$ \endtrivlist}

\title{Extremal and weakly trapped submanifolds in 
Generalized Robertson-Walker spacetimes}

\author{Jos\'e A. S. Pelegr\'in \\[6mm]
Departamento de Did\'actica de la Matem\'atica, \\[0.5mm]
Universidad Granada, 18071, Granada, Spain \\ E-mail\textup{:
\texttt{jpelegrin@ugr.es}} \\[3mm]}

\date{}

\begin{document}

\maketitle

\thispagestyle{empty}

\begin{abstract}
In this article we obtain new rigidity results for spacelike submanifolds of 
arbitrary codimension 
in Generalized Robertson-Walker spacetimes. Namely, 
under appropriate assumptions such as parabolicity we prove by means of some maximum 
principles that they must be contained in a 
spacelike slice. This enables us to characterize extremal and weakly 
trapped submanifolds in these ambient spacetimes.
\end{abstract}
\vspace*{5mm}



\section{Introduction}

The study of spacelike submanifolds has a long and fruitful history in 
General Relativity. Indeed, codimension one spacelike submanifolds 
represent the physical universe that can be measured at a given 
moment of time and they play a crucial role in a variety of results 
that include from the Cauchy problem \cite{CG} to 
Causality Theory \cite{BEE}. Spacelike submanifolds of 
codimension greater than one began to attract scientific interest
when Penrose introduced the notion of trapped surfaces in four 
dimensional spacetimes to study the singularities of a 
spacetime \cite{Pe}. Namely, the existence of a trapped surface 
indicates the presence of a black hole. Thus, spacelike  
submanifolds of codimension two have been key 
to prove some of the most important singularity 
theorems \cite{HP}, analyze gravitational 
collapse and black holes formation \cite{BM} as well as in the
study of the cosmic censorship 
hypothesis \cite{CIM} and the related Penrose inequality \cite{HI}.

Trapped surfaces are usually defined in terms of null expansions. However, 
relating these null expansions to the causal 
orientation of the mean curvature vector 
provides a better characterization of these 
surfaces and allows the extension of the concepts of trapped submanifolds to  
arbitrary codimension \cite{Kr}. Indeed, let 
$\psi: M^k \longrightarrow \overline{M}^{n}$ be a connected immersed 
spacelike submanifold of arbitrary codimension $n-k$ 
in an $n$-dimensional spacetime (i.e., the induced metric on $M$ 
is Riemannian). Denoting by $\overrightarrow{H}$ 
the mean curvature vector field of the immersion, following the 
standard terminology in General Relativity (see \cite{Kr} and 
\cite{MS}), the spacelike submanifold is said to be:

\begin{itemize}
\item \textit{Future (past) trapped} if $\overrightarrow{H}$ is timelike and 
future (past) pointing everywhere on $\psi(M)$.

\item \textit{Marginally future (past) trapped} if $\overrightarrow{H}$ is lightlike 
and future (past) pointing everywhere on $\psi(M)$. 

\item \textit{Weakly future (past) trapped} if $\overrightarrow{H}$ is causal and 
future (past) pointing everywhere on $\psi(M)$.

\item \textit{Extremal} if $\overrightarrow{H} = 0$ on $\psi(M)$.
\end{itemize}

The aforementioned reasons have recently aroused 
the interest in obtaining new rigidity 
results for spacelike submanifolds in a 
wide variety of spacetimes including Lorentzian space forms \cite{CVW, LSV}, 
pp-wave spacetimes \cite{MS, PRR1}, Robertson-Walker spacetimes 
\cite{AnC} and standard static 
spacetimes \cite{CPR, Pe5} among others. 

In this article we will study spacelike submanifolds of arbitrary codimension 
in the family of cosmological models known as Generalized Robertson-Walker 
(GRW) spacetimes. GRW spacetimes are Lorentzian warped products whose negative 
definite base represents a universal time and whose fiber is an 
arbitrary Riemannian manifold (see Section \ref{s2}). These spacetimes 
extend the classical Robertson-Walker models to the case where the fiber 
does not necessarily have constant sectional curvature \cite{A-R-S1}. Since 
GRW spacetimes are not necessarily spatially homogeneous, they 
become suitable cosmological models to describe the universe 
in a more accurate scale \cite{RaS}. 

The first characterization results for spacelike submanifolds in 
these models were obtained for codimension one spacelike submanifolds
in spatially closed GRW spacetimes that satisfied certain energy conditions 
\cite{A-R-S1}. Since spatially open spacetimes provide more accurate models 
of our current universe \cite{Chiu} and do not lead to a violation of 
the holographic principle \cite{Pe2}, these uniqueness 
results for codimension one 
spacelike submanifolds were extended to spatially 
open GRW spacetimes by means of parabolicity \cite{RRS} and 
maximum principles valid for complete Riemannian 
manifolds \cite{LaR, PeR, Pe7}. Indeed, those results 
have been extended to spacelike submanifolds of arbitrary 
codimension in GRW spacetimes by assuming certain  
energy conditions on the spacetime \cite{ACC} as well as 
some bounds on the gradient of the submanifold's 
height function \cite{CLLS} using different maximum principles. 

Our aim in this article is to obtain new rigidity results for 
spacelike submanifolds of arbitrary codimension in GRW spacetimes. In 
order to do so, we will use certain maximum principles 
for parabolic Riemannian manifolds (which include the closed case) 
as well as a maximum principle for complete non-compact Riemannian 
manifolds without imposing any energy condition on the 
ambient spacetime nor a bound on the gradient of the submanifold's 
height function.

This article is organized as follows. In Section \ref{s2} we
define our ambient spacetimes as well as present the 
geometric setup that will be used along the paper. Section \ref{sess} 
is devoted to study the particular case of submanifolds 
which are contained in a spacelike slice of a GRW spacetime, obtaining some 
results that will later help us characterizing spacelike 
submanifolds in more 
general cases. Finally, in Section \ref{smr} we obtain our main rigidity 
results for general spacelike submanifolds of arbitrary codimension which 
state that, under certain conditions, they are contained in a spacelike 
slice. In particular, we obtain in Theorem \ref{teohalf} some 
restrictions on the height function of spacelike 
submanifolds in GRW spacetimes with certain expanding/contracting 
behaviour that allows us to deduce some rigidity and 
non-existence results for the compact case. Moreover, for parabolic 
spacelike submanifolds of arbitrary codimension we obtain our 
main rigidity results under natural assumptions on the behaviour 
of the spacetime (Theorem \ref{teopar}) as well as under a certain 
bound for the mean curvature of the spacelike 
submanifold (Theorem \ref{teoparslab}). Furthermore, by means of a 
maximum principle for complete Riemannian manifolds we are able to 
extend our results for the non-parabolic case (Theorem \ref{teoasy}). As 
a consequence of our main theorems we obtain several results for extremal and 
weakly trapped submanifolds in some well-known GRW spacetimes.

\section{Preliminaries}
\label{s2} 

Let $(F,g_F)$ be an $(n-1)(\geq 2)$-dimensional (connected) Riemannian 
manifold, $I$ an open interval in 
$\mathbb{R}$ and $f$ a positive smooth function defined on $I$. Consider now
 the product manifold 
$\overline{M} = I \times F$ endowed with the Lorentzian metric

\begin{equation}
\label{metrica}
\overline{g} = -\pi^*_{I} (dt^2) +f(\pi_{I})^2 \, \pi_{F}^* (g_F), 
\end{equation}

\noindent where $\pi_{I}$ and $\pi_{F}$ denote the projections onto $I$ and
$F$, respectively. The Lorentzian manifold $(\overline{M}, \overline{g})$ is 
a warped product (in the 
sense of \cite[Chap. 7]{O'N}) with base $(I,-dt^2)$, 
fiber $(F,g_F)$ and warping function $f$. If we endow
$(\overline{M}, \overline{g})$ with the time orientation 
induced by $\partial_t := \partial / \partial t$ 
we can call it, in agreement with the terminology introduced 
in \cite{A-R-S1}, an $n$-dimensional 
Generalized Robertson-Walker (GRW) spacetime and represent it by 
$\overline{M} = I \times_f F$.

In any GRW spacetime there is a distinguished timelike and future pointing
vector field, $\xi: = f({\pi}_I)\partial_t$ that satisfies

\begin{equation}
\label{conexion} 
\overline{\nabla}_X \xi = f'({\pi}_I)\,X
\end{equation}

\noindent for any $X\in \mathfrak{X}(\overline{M})$, where $\overline{\nabla}$
is the Levi-Civita connection of the Lorentzian metric
(\ref{metrica}). Hence, $\xi$ is conformal and its metrically 
equivalent $1$-form is closed.

Let $\psi: M^k \longrightarrow \overline{M}^{n} = I \times_f F$ be an 
immersed spacelike submanifold of arbitrary codimension $ n-k \geq 1$ in an 
$n$-dimensional 
GRW spacetime. Through all this article we will 
assume that our spacelike submanifolds are connected. Since 
the immersion is spacelike, the Lorentzian metric \eqref{metrica} 
induces via $\psi$ a Riemannian metric $g$ on $M$. Let 
us denote by $\overline{\nabla}$ and 
$\nabla$ the Levi-Civita connections of $\overline{M}$ and $M$, 
respectively. The Gauss formula of $M$ in $\overline{M}$ is

\begin{equation}
\label{gaf}
\overline{\nabla}_X Y = \nabla_X Y - \mathrm{II} (X, Y),
\end{equation}

\noindent for every $X, Y \in \mathfrak{X}(M)$, where 
$\mathrm{II}: \mathfrak{X}(M) \times \mathfrak{X}(M) \longrightarrow 
\mathfrak{X}^\perp (M)$ 
denotes the second fundamental form of $M$ given by

\begin{equation}
\label{sff}
\mathrm{II} (X, Y) := - \left( \overline{\nabla}_X Y \right)^\perp.
\end{equation}

Note that we are 
following the usual convention in General Relativity to define 
$\mathrm{II}$ (which is opposite to the usual one in Differential 
Geometry). On the other hand, Weingarten's formula is given by

\begin{equation}
\label{weif}
\overline{\nabla}_X N = A_N X + \nabla_X^\perp N
\end{equation}

\noindent  for $X \in \mathfrak{X}(M)$ and $N \in \mathfrak{X}^\perp (M)$, 
where $A_N$ stands for the shape operator associated to $N$ and $\nabla^\perp$ 
is the normal connection of the submanifold. In addition, we can define the 
mean curvature vector $\overrightarrow{H}$ of $\psi(M)$ by

\begin{equation}
\label{h}
\overrightarrow{H} := \frac{1}{k} \mathrm{trace}(\mathrm{II} ).
\end{equation}

Moreover, we can decompose the vector field $\partial_t$ along  
$\psi(M)$ as

\begin{equation}
\label{dt}
\partial_t = \partial_t^\top + \partial_t^\perp,
\end{equation}

\noindent where $\partial_t^\top \in \mathfrak{X}(M)$ and 
$\partial_t^\perp \in \mathfrak{X}^\perp(M)$ denote, 
respectively, the tangent and normal components of $\partial_t$. Taking 
this into account, we can define the height function 
$\tau := \pi_I \circ \psi$ and obtain that its 
gradient on $\overline{M}$ is given by
$\overline{\nabla} \tau = - \partial_t^\top$. In addition, it is
easy to check that the gradient of $\tau$ on $M$ is

\begin{equation}
\label{grt}
\nabla \tau= - \partial_t^\top,
\end{equation}

Now, using (\ref{grt}) 
and choosing a local orthonormal 
frame $\{E_1, \dots , E_k \}$ on $(M, g)$  we can compute 
the Laplacian of $\tau$ on $M$ (see \cite[Lemma 1]{CLLS}), obtaining

\begin{equation}
\label{laptau1}
\Delta \tau = - \sum_{i=1}^k g\left( \nabla_{E_i} \partial_t^\top , E_i\right) 
=  k \ \overline{g}( \overrightarrow{H}, \partial_t) 
- \left(k + |\nabla \tau |^2 \right) \frac{f'(\tau)}{f(\tau)},
\end{equation}

\noindent where we have used \eqref{conexion} and Gauss formula 
\eqref{gaf}.

Among the spacelike submanifolds of codimension one in a GRW spacetime 
the spacelike slices $\{t_0\} \times F$, $t_0 \in I$ form a remarkable 
family that folitates the whole spacetime. We can easily see that a 
spacelike submanifold $\psi(M)$ is contained in a spacelike slice if and only if 
$\tau$ is constant on $M$. In addition, the spacelike slices are the 
restspaces of the so called comoving observers, which are 
determined by the integral curves of $\partial_t$ \cite{SW}. 

Let us denote by $\Sigma_{t_0} = \{t_0\} \times F$ a spacelike slice 
of a GRW spacetime $\overline{M} = I \times_f F$ for $t_0 \in I$ and 
compute its second fundamental form. To do so, from \eqref{conexion} we 
have that

\begin{equation}
\label{dxpt}
\overline{\nabla}_X \partial_t = \frac{f'(t_0)}{f(t_0)} X,
\end{equation}

\noindent for every $X \in \mathfrak{X}(\Sigma_{t_0})$. Using this and 
\eqref{sff}, the second fundamental form of the spacelike slice 
$\mathrm{II}_{\Sigma_{t_0}}$ is

\begin{equation}
\label{sffsl}
\mathrm{II}_{\Sigma_{t_0}} (X, Y) = 
- \frac{f'(t_0)}{f(t_0)} \overline{g}(X, Y) \partial_t.
\end{equation}

Thus, we can clearly see from \eqref{sffsl} that spacelike slices 
are totally umbilical and their mean curvature vector is given by

\begin{equation}
\label{hsli}
\overrightarrow{H}_{\Sigma_{t_0}} = - \frac{f'(t_0)}{f(t_0)} \partial_t.
\end{equation}

Therefore, a spacelike slice is totally geodesic if and only if 
$f'(t_0) = 0$. Moreover, this mean curvature has a physical interpretation in terms 
of the spacetime behaviour measured by the comoving observers. Indeed, 
for each $p \in \Sigma_{t_0}$, computing the 
divergence in $\overline{M}$ of $\partial_t$ at $p$ we obtain

\begin{equation}
\label{diin}
\overline{\mathrm{div}}(\partial_t)_p = (n-1) \frac{f'(t_0)}{f(t_0)}.
\end{equation}

Hence, the sign of $f'(t)$ determines wether the
comoving observers measure that the are spreading 
out or coming together. In particular, if $f'(t) \leq 0$ 
(resp., $f'(t) \geq 0$) we will say that the spacetime is non-expanding 
(resp., non-contracting).

\subsection{Parabolic Riemannian manifolds}

A complete (non-compact) Riemannian manifold is called 
parabolic if the only superharmonic 
functions bounded from below that it admits are the constants (see \cite{Ka} 
for example). From a physical standpoint,  
parabolic Riemannian manifolds are those 
where the Brownian motion is recurrent \cite{Gr}. This property has 
an interesting
physical interpretation: any particle in the Riemannian manifold
following a Brownian motion will pass
through any open subset at some arbitrarily large moment of time.

From a mathematical perspective, the parabolicity of a Riemannian surface is 
closely related to its Gaussian curvature. In fact, 
complete spacelike surfaces with non-negative Gaussian curvature are 
parabolic \cite{H}. For Riemannian manifolds of arbitrary 
dimension, despite not being a clear relation between parabolicity and 
the sectional curvature, there are sufficient conditions 
to ensure their 
parabolicity based 
on the geodesic balls' volume growth \cite{AMR, Gr}. For example, 
if a spacelike submanifold $\psi: M \longrightarrow \overline{M}$ is 
complete, we can ensure its parabolicity if 
the geodesic balls $B_R$ with a fixed origin $o \in M$ 
satisfy

$$ \frac{R}{\mathrm{vol}(B_R)} \not\in L^1(+\infty),$$

\noindent or

$$\frac{1}{\mathrm{vol}(\partial B_R)} \not\in L^1(+\infty).$$
 
In addition, another crucial property about parabolicity is that it is invariant 
under quasi-isometries (\cite{Ka}, \cite[Cor. 5.3]{Gr}). Recall 
that a quasi-isometry between two Riemannian 
manifolds $(M_1, g_1)$ and $(M_2, g_2)$ is
a global diffeomorphism $\varphi$ from $M_1$ onto $M_2$ such 
that for some constant $c \geq 1$ satisfies

\begin{equation*}
c^{-1} \ g_1(v, v) \leq g_2(d\varphi(v), d\varphi(v)) \leq c \ g_1(v, v)
\end{equation*}

\noindent for all $v \in T_pM_1$, $p \in M_1$. Using this 
property, in \cite{RRS} the authors 
obtain sufficient conditions to ensure that the parabolicity of the 
fiber of a GRW spacetime is inherited by every complete 
spacelike hypersurface. This provides a way to 
guarantee the parabolicity of complete
spacelike hypersurfaces in a GRW spacetime. Therefore, 
we can ensure the parabolicity of a submanifold of arbitrary 
codimension in a GRW spacetime if it can be contained in 
a parabolic spacelike hypersurface.

\section{Submanifolds contained in spacelike slices}
\label{sess}

Let $\phi: M^k \longrightarrow F^{n-1}$ be a $k$-dimensional immersed submanifold in
the Riemannian $(n-1)$-dimensional manifold $(F, g_F)$ 
and denote by $g$ the Riemannian metric 
induced on $M$ by $\phi$, i.e., $g = \phi^* (g_F)$. Consider for a 
fixed $t_0 \in I$ the map 
$\phi_{t_0}: M \longrightarrow \overline{M} = I \times_f F$ into the 
GRW spacetime $(\overline{M} = I \times_f F, \overline{g})$ 
given by 

$$ \phi_{t_0} (p) = (t_0, \phi(p)), \ \text{for every} \ p \in M.$$

We see that $\phi_{t_0}$ is an immersion 
of $M$ into $\overline{M}$ which is contained in the spacelike slice 
$\Sigma_{t_0} = \{t_0\} \times F$. Moreover, 
the induced metric on $M$ via $\phi_{t_0}$ is 

$$g_{t_0} = \phi_{t_0}^* (\overline{g}) = \phi^* (g_F) = g.$$

Conversely, given an immersion 
$\psi: M^k \longrightarrow \overline{M}^{n} = I \times_f F^{n-1}$ 
which is contained in a spacelike slice $\Sigma_{t_0} = \{t_0\} \times F$, 
the projection $\phi = \pi_F \circ \psi : M \longrightarrow F$ defines 
an immersion for which 
$\psi(p) = (t_0, \phi(p)) = \phi_{t_0} (p)$. Indeed, 
$\psi: M \longrightarrow \overline{M} = I \times_f F$ 
is an embedding if and only if $\phi: M \longrightarrow F$ is an embedding.

Therefore, we can relate the extrinsic geometry of the submanifold 
$\psi: M \longrightarrow \overline{M} = I \times_f F$ contained in 
the spacelike slice $\Sigma_{t_0} = \{t_0\} \times F$ with 
the extrinsic geometry of $\phi: M \longrightarrow F$. In order to do so, 
let us consider a local orthonormal frame $\{e_1, \dots, e_k\}$ of the 
tangent space to $M$ in $F$ and 
a local orthonormal frame $\{u_{k+1}, \dots, u_{n-1} \}$ of the normal 
vector space to $M$ in $F$. Now, we can easily see that 
$\left\{ E_1 = \frac{e_1}{f(t_0)}, \dots, E_k = \frac{e_k}{f(t_0)} \right\}$ 
and $\left\{ U_{k+1} = \frac{u_{k+1}}{f(t_0)}, \dots, 
U_{n-1} = \frac{u_{n-1}}{f(t_0)}, 
\partial_t \right\}$ define a local orthonormal frame of the tangent and 
normal spaces to $M$ in $\overline{M}$, respectively. 

Therefore, the second fundamental form $\mathrm{II}_\psi$ of the immersion $\psi$ 
can be written as

\begin{equation}
\label{sffa}
\mathrm{II}_\psi (X, Y) = \sum_{i=k+1}^{n-1} 
\overline{g}(\mathrm{II}_\psi(X, Y), U_i) U_i - 
\overline{g}\left(\mathrm{II}_\psi(X, Y), \partial_t \right) \partial_t
\end{equation}
 
\noindent for every $X, Y \in \mathfrak{X}(M)$. Hence, the mean curvature 
vector field of $\psi$ is 

\begin{eqnarray}
\label{Hpsi}
\overrightarrow{H} &=& \frac{1}{k} \mathrm{trace}(\mathrm{II}_\psi) = 
\frac{1}{k} \sum_{i=1}^k \mathrm{II}_\psi (E_i, E_i) = 
- \frac{1}{k} \sum_{i=1}^k \left( \overline{\nabla}_{E_i} E_i \right)^\perp 
\nonumber \\
&=& - \frac{1}{k} \sum_{j=k+1}^{n-1} \sum_{i=1}^k 
\overline{g}\left( \overline{\nabla}_{E_i} E_i, U_j \right) U_j 
+ \frac{1}{k} \sum_{i=1}^k \overline{g}\left( \overline{\nabla}_{E_i} E_i, 
 \partial_t \right) \partial_t.
\end{eqnarray}

Using Gauss formula \eqref{gaf}, the relation between the frames that we have 
previously defined we can compute the first term on the RHS 
of \eqref{Hpsi}, obtaining

\begin{gather}
- \frac{1}{k} \sum_{j=k+1}^{n-1} \sum_{i=1}^k 
\overline{g}\left( \overline{\nabla}_{E_i} E_i, U_j \right) U_j = 
- \frac{1}{k} \sum_{j=k+1}^{n-1} \sum_{i=1}^k 
\overline{g}\left( \nabla_{E_i}^F E_i, U_j \right) U_j \nonumber \\
+ \frac{1}{k} \sum_{j=k+1}^{n-1} \sum_{i=1}^k 
\overline{g}\left( \mathrm{II}_{\Sigma_{t_0}} (E_i, E_i), U_j \right) U_j =  
- \frac{1}{k} \frac{1}{f(t_0)^2} \sum_{j=k+1}^{n-1} \sum_{i=1}^k 
g_F \left( \nabla_{e_i}^F e_i, u_j \right) u_j \nonumber \\
= \frac{1}{k} \frac{1}{f(t_0)^2} \mathrm{trace}( \mathrm{II}_\phi) = 
\frac{\overrightarrow{h}}{f(t_0)^2}, \label{Hpsi2}
\end{gather}

\noindent where we have used \eqref{sffsl} and the orthogonality 
of the chosen frame of the normal space to $M$ in $\overline{M}$ 
to obtain that the term 
involving $\mathrm{II}_{\Sigma_{t_0}}$ identically vanishes. Also, note 
that $\nabla^F$ denotes the Levi-Civita connection on $(F, g_F)$, 
$ \mathrm{II}_\phi$ is the second fundamental form of $\phi$ and 
$\overrightarrow{h}$ stands for the mean curvature vector of $\phi$, 
given by

\begin{equation}
\label{hphi}
\overrightarrow{h} = \frac{1}{k} \mathrm{trace}( \mathrm{II}_\phi).
\end{equation}

Furthermore, the second term on the RHS of \eqref{Hpsi} can be computed using 
\eqref{dxpt}, yielding

\begin{equation}
\label{Hpsi3}
\frac{1}{k} \sum_{i=1}^k \overline{g}\left( \overline{\nabla}_{E_i} E_i, 
 \partial_t \right) \partial_t = 
 - \frac{1}{k} \sum_{i=1}^k \overline{g}\left( E_i, 
 \overline{\nabla}_{E_i} \partial_t \right) \partial_t = 
 - \frac{f'(t_0)}{f(t_0)} \partial_t.
\end{equation}

Thus, inserting \eqref{Hpsi2} and \eqref{Hpsi3} in \eqref{Hpsi} we conclude that

\begin{equation}
\label{Hpsi4}
\overrightarrow{H} = \frac{\overrightarrow{h}}{f(t_0)^2} 
- \frac{f'(t_0)}{f(t_0)} \partial_t.
\end{equation}

From \eqref{Hpsi4} we can directly obtain the following result.

\begin{prop}
\label{profpsli}
Let $\psi: M \longrightarrow \overline{M} = I \times_f F$ be a submanifold  
contained in a spacelike slice $\{t_0\} \times F$, 
$t_0 \in I$ of a GRW spacetime with causal mean curvature 
vector field $\overrightarrow{H}$. Then, $\psi(M)$ is

\begin{itemize}
\item Future pointing iff $f'(t_0) < 0$.

\item Past pointing iff $f'(t_0) > 0$.
\end{itemize}

\end{prop}

Moreover, taking the norm with respect to $\overline{g}$ in \eqref{Hpsi4} we 
have

\begin{equation}
\label{Hpsinorm}
\overline{g}(\overrightarrow{H}, \overrightarrow{H})  = 
\frac{|\overrightarrow{h}|^2_F - f'(t_0)^2}{f(t_0)^2},
\end{equation}

\noindent where $|\overrightarrow{h}|_F$ denotes the norm of $\overrightarrow{h}$
with the metric $g_F$. This enables us to obtain the next characterization 
result for submanifolds contained in spacelike slices (compare with \cite{FHO}).

\begin{prop}
\label{proslice}

Let $\psi: M \longrightarrow \overline{M} = I \times_f F$ be a submanifold  
contained in a spacelike slice $\{t_0\} \times F$, 
$t_0 \in I$ of a GRW spacetime. Then, $\psi(M)$ is

\begin{itemize}

\item trapped iff $|\overrightarrow{h}|_F < |f'(t_0)|$.

\item marginally trapped iff $|\overrightarrow{h}|_F = |f'(t_0)|$ 
and $f'(t_0) \neq 0$.

\item weakly trapped iff $|\overrightarrow{h}|_F \leq |f'(t_0)|$ 
and $f'(t_0) \neq 0$.

\item extremal iff $|\overrightarrow{h}|_F = |f'(t_0)| = 0$.

\end{itemize}

\end{prop}

\section{Main results}
\label{smr} 

In this section we will focus on the 
general case of spacelike submanifolds of arbitrary 
codimension which are not necessarily contained in a spacelike slice 
of a GRW spacetime. In particular, we obtain sufficient conditions 
to guarantee that the submanifold is contained in a spacelike slice, which 
combined with the results of the previous section will allow us to provide new 
rigidity and characterization results.

\subsection{Rigidity results for monotonic warping function}

As a first result, we are able to prove the next theorem.

\begin{teor}
\label{teohalf}

Let $\psi: M \longrightarrow \overline{M} = I \times_f F$ be 
a spacelike submanifold in a GRW spacetime.

\begin{enumerate}[label=(\roman*)]

\item If $\overline{g}(\overrightarrow{H}, \partial_t) < 0$ and $f'(t) \geq 0$
(resp., $\overline{g}(\overrightarrow{H}, \partial_t) \leq 0$ and $f'(t) > 0$), then 
the height function $\tau$ has no local minimum. 

\item If $\overline{g}(\overrightarrow{H}, \partial_t) > 0$ and $f'(t) \leq 0$ 
(resp., $\overline{g}(\overrightarrow{H}, \partial_t) \geq 0$ and $f'(t) < 0$), then 
the height function $\tau$ has no local maximum. 

\end{enumerate}

\end{teor}

\begin{demo}
Reasoning by contradiction, let us assume that there exists a point $p \in M$ 
where $\tau$ attains a local minimum. Hence, $\nabla \tau (p) = 0$ and 
$\Delta \tau (p) \geq 0$. Nevertheless, if 
the assumptions in the first statement hold, 
\eqref{laptau1} yields

$$\Delta \tau (p) =  k \ \overline{g}( \overrightarrow{H}, \partial_t)(p)
- k \frac{f'(\tau)}{f(\tau)}(p) < 0,$$

\noindent obtaining a contradiction. We can proceed analogously to prove 
the second statement.
\end{demo}

As a consequence of Theorem \ref{teohalf}, we deduce the next non-existence 
result for closed (i.e., compact without boundary) 
weakly trapped submanifolds of arbitrary 
codimension (previously obtained in \cite[Cor. 5.2]{ACC}).

\begin{coro}
\label{corwt}
Let $\overline{M} = I \times_f F$ be a GRW spacetime.

\begin{enumerate}[label=(\roman*)]

\item If $\overline{M}$ is non-contracting, then there are no closed 
weakly future trapped submanifolds in $\overline{M}$.

\item If $\overline{M}$ is non-expanding, then there are no closed 
weakly past trapped submanifolds in $\overline{M}$.

\end{enumerate}

\end{coro}

Moreover, from Theorem \ref{teohalf} we can also 
obtain the following corollary for the extremal case, 
which can be seen as an extension to arbitrary codimension 
of \cite[Prop. 4.1]{A-R-S1}.

\begin{coro}
\label{cormin}

There are no closed extremal submanifolds in a strictly expanding (resp., 
strictly contracting) GRW spacetime.
\end{coro}

We are now in a position to obtain new rigidity results for  
spacelike submanifolds of arbitrary codimension immersed in a GRW spacetime 
which are not necessarily closed. In particular, assuming the 
parabolicity of the submanifold we obtain the following theorem. 

\begin{teor}
\label{teopar}

Let $\psi: M \longrightarrow \overline{M} = I \times_f F$ be 
a parabolic spacelike submanifold in a GRW spacetime.

\begin{enumerate}[label=(\roman*)]

\item If $\overline{g}(\overrightarrow{H}, \partial_t) \leq 0$, 
$f'(\tau) \geq 0$ and $\inf_M \tau > -\infty$, then $\psi(M)$ is 
contained in a totally geodesic spacelike slice.

\item If $\overline{g}(\overrightarrow{H}, \partial_t) \geq 0$, 
$f'(\tau) \leq 0$ and $\sup_M \tau < +\infty$, then $\psi(M)$ is 
contained in a totally geodesic spacelike slice.

\end{enumerate}
\end{teor}

\begin{demo}

If the assumptions in the first statement hold, from \eqref{laptau1}, 
the Laplacian of the height function verifies

$$\Delta \tau =  k \ \overline{g}( \overrightarrow{H}, \partial_t) 
- \left(k + |\nabla \tau |^2 \right) \frac{f'(\tau)}{f(\tau)} \leq 0.$$

Hence, $\tau$ is a superharmonic function bounded from below in a parabolic 
Riemannian manifold. Therefore, it must be constant. The proof 
of the second statement is analogous.
\end{demo}

Consequently, from this theorem we deduce the next non-existence result for 
weakly trapped parabolic submanifolds in a GRW spacetime, which extends 
Corollary \ref{corwt}.

\begin{coro}
\label{corpar1}

Let $\overline{M} = I \times_f F$ be a GRW spacetime.

\begin{enumerate}[label=(\roman*)]

\item If $\overline{M}$ is non-contracting, then there are no weakly 
future trapped parabolic submanifolds bounded away from past 
infinity.

\item If $\overline{M}$ is non-expanding, then there are no weakly 
past trapped parabolic submanifolds bounded away from future 
infinity.

\end{enumerate}

\end{coro}

In addition, for the extremal case we can combine Theorem \ref{teopar} and 
Proposition \ref{proslice} to obtain the following extension of 
Corollary \ref{cormin} to the parabolic case.

\begin{coro}
\label{coromin}

Let $\overline{M} = I \times_f F$ be a GRW spacetime.

\begin{enumerate}[label=(\roman*)]

\item If $\overline{M}$ is non-contracting, then every extremal 
parabolic submanifold bounded away from past 
infinity is contained as a minimal submanifold in 
a totally geodesic spacelike slice.

\item If $\overline{M}$ is non-expanding, then every extremal 
parabolic submanifold bounded away from future
infinity is contained as a minimal submanifold in 
a totally geodesic spacelike slice.

\end{enumerate}

\end{coro}

Furthermore, from Theorem \ref{teopar} we 
can also obtain the following 
corollary for the particular case where the 
warping function $f(t)$ is constant, having a Lorentzian product 
spacetime $\mathbb{R} \times F$ where the timelike vector field 
$\partial_t$ is Killing. Lorentzian products include spacetimes 
such as Lorentz-Minkowski spacetime $\mathbb{L}^n$ and 
Einstein static universe $\mathbb{R} \times \mathbb{S}^{n-1}$. Indeed, we 
obtain in these models the next rigidity result 
for extremal and weakly trapped parabolic submanifolds contained in 
a slab (i.e., between two spacelike slices), 
which in particular include the closed ones. Therefore, 
this corollary generalizes the result obtained for 
the extremal closed case in \cite[Cor. 8]{AER} 
(compare also with \cite[Thm. 10]{Pe5}).

\begin{coro}
\label{coropro}

In a Lorentzian product spacetime $\mathbb{R} \times F^{n-1}$ there are no 
weakly trapped parabolic submanifolds contained in a slab. Moreover, the only 
extremal parabolic submanifolds bounded 
away from either future or past infinity in $\mathbb{R} \times F^{n-1}$ 
are the spacelike slices' minimal 
submanifolds. 
\end{coro}

\begin{rem}
\normalfont

Notice that the parabolicity assumption in Theorem \ref{teopar} cannot be 
weakened to completeness in general. Indeed, using the upper half-plane 
model for the hyperbolic plane, i.e., 
$\mathbb{H}^2 = \{ (x_1, x_2) \in \mathbb{R}^2, x_2 >0\}$ endowed with the 
metric $g_{\mathbb{H}^2} = \frac{1}{x_2^2}(x_1^2 + x_2^2)$,  
omitting the parabolicity assumption in Theorem \ref{teopar} we obtain that in 
the Lorentzian product spacetime 
$(\mathbb{R}\times \mathbb{H}^2, -dt^2 + g_{\mathbb{H}^2})$ 
the function 
$u(x_1, x_2) = \frac{1}{4} \log(x_1^2 + x_2^2)$, $(x_1, x_2) \in \mathbb{H}^2$
defines a non-trivial complete spacelike graph with zero mean curvature bounded 
away from past infinity (see \cite{Alb}).

\end{rem}

\subsection{Rigidity results for bounded mean curvature}

Assuming that the timelike component 
of the mean curvature vector of the spacelike submanifold 
satisfies certain bound and that the spacelike submanifold is 
contained in a slab 
we can extend our results to GRW spacetimes 
where the warping function does not necessarily have a monotonic behaviour, 
obtaining the next theorem.

\begin{teor}
\label{teoparslab}

Let $\psi: M \longrightarrow \overline{M} = I \times_f F$ be 
a parabolic spacelike submanifold in a GRW spacetime 
contained in a slab. If

\begin{equation}
\label{hf1}
\overline{g}(\overrightarrow{H}, \partial_t) \leq \frac{f'(\tau)}{f(\tau)}
\end{equation}

\noindent or

\begin{equation}
\label{hf2}
\overline{g}(\overrightarrow{H}, \partial_t) \geq \frac{f'(\tau)}{f(\tau)},
\end{equation}

\noindent then $\psi(M)$ is 
contained in a spacelike slice.

\end{teor}

\begin{demo}

Since $\psi(M)$ is in a slab, we have $\tau \in (c, d)$ for certain 
$c, d \in \mathbb{R}$. Thus, we can define the 
primitive function of $f$ in $(c, d)$, given by

$$\mathcal{F}(t) = \int_c^t f(s) ds, \ t \in (c, d).$$

Hence, $\mathcal{F}(\tau):= \mathcal{F} \circ \tau$ is a positive 
function on $M$ which is also bounded from above since $\psi(M)$ is in a slab. Now,
from \eqref{laptau1} we deduce

\begin{equation}
\label{laF}
\Delta \mathcal{F}(\tau) = 
 k f(\tau) \overline{g}( \overrightarrow{H}, \partial_t) - k f'(\tau).
\end{equation}

If either \eqref{hf1} or \eqref{hf2} hold, $\mathcal{F}(\tau)$ is a 
bounded function with signed Laplacian in 
a parabolic Riemannian manifold, so it must be constant. Therefore, 
$\nabla \mathcal{F}(\tau) = f(\tau) \nabla \tau = 0$ and 
the result follows.
\end{demo}

\begin{rem}
\label{rema3}
\normalfont
Note that \eqref{hf1} 
and \eqref{hf2} relate at each point $p \in M$ of the spacelike submanifold 
the timelike component of the mean curvature 
vector field of $\psi(M)$ with minus the mean curvature function 
of the spacelike slice 
that passes through that point \eqref{hsli}. Therefore, these are not comparison 
assumptions between extrinsic properties of two manifolds, since the RHS
corresponds to a spacelike slice that changes at each point. Similar 
bounds have been previously used to obtain characterization 
results for the codimension one case \cite{CRR, PRR5}.

In addition, if we weaken the assumptions in Theorem \ref{teoparslab} we 
find examples of spacelike submanifolds in GRW spacetimes that 
are not contained in a spacelike slice. In particular, if we 
omit the assumption of being in a slab, we obtain in the 
three-dimensional Lorentz-Minkowski spacetime $\mathbb{L}^3$ 
that the non-horizontal spacelike planes satisfy the remaining
assumptions but are not spacelike slices (see also \cite{HO} for examples 
of boost invariant marginally trapped surfaces in $\mathbb{L}^4$). More 
examples of 
marginally trapped surfaces in Robertson-Walker spacetimes that 
are not contained in a spacelike slice can be found 
by means of \cite[Thm. 2]{AnC}.

\end{rem}

Moreover, for the well-known Robertson Walker models given by the 
steady state and Einstein-de Sitter spacetimes of arbitrary dimension 
we can proceed as in the proof of Theorem \ref{teoparslab} to deduce the 
following non-existence results for extremal weakly trapped submanifolds.

\begin{prop}
\label{coross}

In the steady state spacetime $\mathbb{R} \times_{e^t} \mathbb{R}^{n-1}$ 
there are no weakly future trapped nor extremal parabolic submanifolds.
\end{prop}

\begin{demo}
If we compute the Laplacian of the warping function restricted to $M$, $e^\tau$, 
using \eqref{laptau1} we obtain 

$$\Delta e^\tau = k e^\tau \overline{g}(\overrightarrow{H}, \partial_t) - k.$$

Thus, if $\psi(M)$ was weakly future trapped or extremal, $e^\tau$ 
would be a non-negative 
superharmonic function on a parabolic Riemannian manifold. Therefore, it 
would be constant and the spacelike submanifold 
would be contained in a spacelike slice, reaching a contradiction 
in both cases due to Propositions \ref{profpsli} and \ref{proslice}.
\end{demo}

\begin{prop}
\label{coroeds}

In Einstein-de Sitter spacetime $\mathbb{R}^+ \times_{t^{2/3}} \mathbb{R}^{n-1}$ 
there are no weakly future trapped nor extremal parabolic submanifolds.
\end{prop}

\begin{demo}
In this case, since $I = \mathbb{R}^+$ we have from \eqref{laptau1}

$$\Delta \tau = k \overline{g}(\overrightarrow{H}, \partial_t) 
- \frac{2}{3 \tau} \left(k + |\nabla \tau |^2 \right).$$

Hence, if $\psi(M)$ was weakly future trapped or extremal, $\tau$ 
would be a non-negative 
superharmonic function on a parabolic Riemannian manifold, so it 
would be constant. Thus, the spacelike submanifold 
would be contained in a spacelike slice, reaching a contradiction 
in both cases due to Propositions \ref{profpsli} and \ref{proslice}.
\end{demo}

To conclude, we will deal with complete spacelike submanifolds which are not 
necessarily parabolic nor contained in a slab
by means of a maximum principle obtained in \cite{ACN}. 
 
\begin{teor}
\label{teoasy}
Let $\psi: M \longrightarrow \overline{M} = I \times_f F$ be 
a complete non-compact oriented spacelike submanifold 
in a GRW spacetime whose warping function satisfies $f(t) \in L^1(I)$.

\begin{enumerate}[label=(\roman*)]

\item If $\overline{g}(\overrightarrow{H}, \partial_t) \leq 
\frac{f'(\tau)}{f(\tau)}$, $\psi(M)$ is below a 
spacelike slice $\{t_0\} \times F$, $t_0 \in I$ and 
asymptotic to it at infinity, then $\psi(M)$ is contained in $\{t_0\} \times F$.

\item If $\overline{g}(\overrightarrow{H}, \partial_t) \geq 
\frac{f'(\tau)}{f(\tau)}$, 
$\psi(M)$ is above a spacelike slice $\{t_0\} \times F$, $t_0 \in I$ and 
asymptotic to it at infinity, then $\psi(M)$ is contained in $\{t_0\} \times F$.

\end{enumerate}

\end{teor}

\begin{demo}
Since the warping function is integrable in the whole interval $I = (a, b)$ 
we can define as in the proof of Theorem \ref{teoparslab} a primitive 
function

$$\mathcal{F}(t) = \int_a^t f(s) ds, \ t \in (a, b)$$

\noindent and denote $\mathcal{F}(\tau) := \mathcal{F} \circ \tau$. Hence, 
under our assumptions in \textit{(i)} the Laplacian of $-\mathcal{F}(\tau)$
satisfies

$$\Delta (-\mathcal{F}(\tau)) = 
-k f(\tau) \overline{g}( \overrightarrow{H}, \partial_t) 
+ k f'(\tau) \geq 0.$$

Consider now on $M$ the function 
$v := t_0 - \tau$. Since $\psi(M)$ is 
below $\{t_0\} \times F$ and asymptotic to it at infinity, this 
function verifies $v \geq 0$ on $M$ and 
$\lim_{r(p) \rightarrow + \infty} v(p) = 0$, where $r(p) = d(p, o)$ denotes 
the Riemannian distance function on $M$ measured from a fixed origin 
$o \in M$. In addition, $g(\nabla (-\mathcal{F}(\tau)), \nabla v) = 
f(\tau) |\nabla \tau|^2 \geq 0$.

Reasoning by contradiction, if we suppose that 
$\psi(M) \not\subseteq \{t_0\} \times F$, then for some $p \in M$ we 
have $v(p) > 0$, so that $v \not\equiv 0$ on $M$. Therefore, we can 
apply \cite[Thm. 2.2]{ACN} to conclude that

$$ 0 \equiv g(\nabla (-\mathcal{F}(\tau), \nabla v) = f(\tau) |\nabla \tau|^2.$$

Since $\psi(M)$ is asymptotic to $\{t_0\} \times F$ we obtain that $v \equiv 0$
on $M$, reaching a contradiction. The second statement is proved analogously 
computing the gradient and Laplacian 
of $\mathcal{F}(\tau)$ and using the function $u:= \tau - t_0$ 
instead of $v$.
\end{demo}

\begin{rem}
\normalfont

Note that the integrability assumption $f \in L^1(I)$ in Theorem \ref{teoasy} 
is needed to ensure the existence of a primitive of the warping function 
restricted to the submanifold $M$, $\mathcal{F}(\tau)$. Thus, 
it can be omitted if $\psi(M)$ lies in a slab defining $\mathcal{F}(\tau)$ 
as in the proof of Theorem \ref{teoparslab}.
\end{rem}

\section*{Acknowledgements}

The author is supported by Spanish MINECO and ERDF project MTM2016-78807-C2-1-P. The 
author would like to thank the referee for his deep reading and 
valuable suggestions.

\end{document}